\documentclass[11pt,twoside]{article}
\usepackage{asp2010}
\usepackage{epsf}
\usepackage{graphics}

\resetcounters

\def\spose#1{\hbox to 0pt{#1\hss}}
\def\ltsimm{\mathrel{\spose{\lower 3pt\hbox{$\sim$}}
        \raise 2.0pt\hbox{$<$}}}
\def\gtsimm{\mathrel{\spose{\lower 3pt\hbox{$\sim$}}
        \raise 2.0pt\hbox{$>$}}}
\def\Mdot{\hbox{${\dot M}$}}

\def\km{{\rm\thinspace km}}
\def\cm{{\rm\thinspace cm}}
\def\s{{\rm\thinspace s}}
\def\yr{{\rm\thinspace yr}}
\def\g{{\rm\thinspace g}}

\def\kmps{\hbox{${\rm\km\s^{-1}\,}$}}

\def\erg{{\rm\thinspace erg}}
\def\Hz{{\rm\thinspace Hz}}

\def\ster{{\rm\thinspace ster}}
\def\ergps{\hbox{${\rm\erg\s^{-1}\,}$}}
\def\Msol{\hbox{${\rm\thinspace M_{\odot}}$}}

\def\Msolpyr{\hbox{${\rm\Msol\yr^{-1}\,}$}}
\def\pcm{\hbox{${\rm\cm^{-1}\,}$}}
\def\pcm3{\hbox{${\rm\cm^{-3}\,}$}}
\def\ergpscm3Hz{\hbox{${\rm\ergps\cm^{-3}\Hz^{-1}\,}$}}
\def\ergpscm3Hzster{\hbox{${\rm\ergps\cm^{-3}\Hz^{-1}\ster^{-1}\,}$}}
\def\gpcm3{\hbox{${\rm\g\cm^{-3}\,}$}}
\def\ergpcm2{\hbox{${\rm\erg\cm^{-2}\,}$}}
\def\ergpcm3{\hbox{${\rm\erg\cm^{-3}\,}$}}

\begin{document}

\title{Feedback from Winds and Supernovae in Massive Stellar Clusters}
\author{J. M. Pittard and H. Rogers}
\affil{School of Physics and Astronomy, The University of Leeds, Leeds, LS2 9JT, UK}

\begin{abstract}
We simulate the effects of massive star feedback, via winds and
SNe, on inhomogeneous molecular material left over from the formation
of a massive stellar cluster. We use 3D hydrodynamic models
with a temperature dependent average particle mass to model the
separate molecular, atomic, and ionized phases. We find that the
winds blow out of the molecular clump along low-density channels, and
gradually ablate denser material into these. However, the dense molecular gas
is surprisingly long-lived and is not immediately affected by the
first star in the cluster exploding.
\end{abstract}

\section{Introduction}
Massive stars dramatically affect the interstellar medium surrounding
them, through their powerful winds, radiation fields and terminal
explosions. Their influence can extend to extra-galactic
scales, with groups of massive stars able to pressurize kpc-scale
superbubbles which can vent into the inter-galactic medium. This
``feedback'' can alter the evolution of the host galaxy. The current
consensus is that momentum and energy from stellar winds and
supernovae plays a key role in suppressing star formation in lower
mass dark matter haloes, thus explaining the lack of faint galaxies
relative to bright galaxies \citep{White:1978}.

However, the degree to which these stellar inputs couple to the
clumpy, inhomogeneous molecular clouds which initially surround a
massive stellar cluster is exceedingly ill-determined. Even the 
dominant feedback process continues to be debated
\citep[e.g.][]{Yorke:1989,Draine:1991,Matzner:2002,Wang:2010,Lopez:2011}.
Chandra X-ray and Spitzer infrared observations of clusters of
young stars have shown that the surrounding cold molecular
can sometimes confine the hot gas created by the stars
\citep[e.g.][]{Townsley:2006}, yet in other clusters the
cold clouds appear to be shaped and removed by the hot gas \citep[see,
e.g.,][]{Townsley:2003}. Most simulations of stellar feedback continue
to adopt a uniform ambient medium
\citep[e.g.][]{Freyer:2003,Toala:2011} - those that do not generally
focus on the impact of massive star feedback on local star formation
\citep[e.g.][]{Peters:2010}, rather than larger-scale effects.

There are several competing models for stellar wind
feedback. In the model of \citet{Castor:1975} - see also
\citet{Dyson:1972} and \citet{Weaver:1977} - the shocked stellar wind
is confined by a cool shell of swept-up interstellar gas. In contrast,
\citet{Chevalier:1985} ignored any surrounding material and simply
assumed a steady-state wind extending to infinity. The temperatures
and pressures of the resulting structures are significantly
different. More recently, \citet{Harper-Clark:2009} have examined a
third scenario in which breaks in the swept-up shell allow some of the
hot high-pressure gas in the bubble interior to leak out. This
situation is expected to occur when the surrounding environment is
highly structured, such as in massive star forming regions. A recent
analysis of the pressures of various components inside the giant HII
region 30\,Doradus implied that such leakage is occuring
\citep{Lopez:2011}.  However, the wider picture is not so
clear. Models of the collisions of multiple stellar winds within a star
cluster have also been created
\citep[e.g.][]{Rodriguez-Gonzalez:2008,Reyes-Iturbide:2009}, though
key processes including particle acceleration at
the multiple shocks within the cluster
\citep[e.g.][]{Domingo-Santamaria:2006} and mass-loading from the cold clouds
embedded within it \citep[e.g.][]{Tsivilev:2002,Bruhweiler:2010} have
been neglected to date.

The effects of the ionizing radiation from massive stars were studied
by \citet{Tenorio-Tagle:1979}, who found that an O-star near the edge
of a molecular cloud creates a high pressure H{\sc II} region around
itself which can burst out of the cloud as a ``champagne''
flow. \citet{Whitworth:1979} and \citet{Bodenheimer:1979} showed that
champagne flows could efficiently disrupt the molecular cloud, though
\citet{Mazurek:1980} and \citet{Yorke:1989} argued that the dispersal
of clouds is dramatically reduced if the clouds are in free-fall and
the ionizing sources are located near their centres. More recent work
has shown that ionization feedback into a highly inhomogeneous medium
may not be very effective
\citep[e.g.][]{MacLow:2007,Peters:2010,Krumholz:2010,Dale:2011}.

The effect of a supernova explosion on a molecular clump was
investigated by \citet{Tenorio-Tagle:1985}, who concluded that a
single supernova could disrupt $\sim 10^{4}\Msol$ of molecular cloud
material. However, it is not clear how this inference might change if
the surrounding material is clumpy.

\section{Numerical Models}
We have constructed 3D hydrodynamical models of the
collective influence of stellar winds and supernovae from the stars in
a massive stellar cluster on the molecular, atomic, and ionized gas
within and external to the cluster-forming GMC clump. A key difference
with some earlier works is that we 
do not assume that the initial
ambient medium is uniform and stationary. Instead, for our initial
conditions we adopt the simulation results of
\citet{Vazquez-Semadeni:2008} of turbulent and clumpy molecular clouds
(specifically model Ms24J6). 
We scale these results to create a GMC clump of radius 5\,pc and mass
5000\,\Msol. The clump initially has a uniform temperature of about
10\,K, and is in rough pressure equilibrium with a surrounding uniform
medium of density $3.33\times10^{-25}\,{\rm g\,cm^{-3}}$ and
temperature 8000\,K.  The mechanical feedback from the stellar cluster
is assumed to be dominated by three $40\,\Msol$ stars situated at the centre of
the clump which collectively drive a cluster wind with a mass-loss rate
$\Mdot_{\rm cl} = 1.5\times10^{-6}\,\Msolpyr$ and velocity $v_{\rm cl}
= 2000 \kmps$.  After 4.0\,Myr the cluster wind is assumed to be
dominated by red-super-giant material, and has $\Mdot_{\rm cl} =
3\times10^{-4}\,\Msolpyr$ and $v_{\rm cl} = 50 \kmps$. The cluster
wind transitions to a Wolf-Rayet dominated phase after
4.2\,Myr with $\Mdot_{\rm cl} = 6\times10^{-5} \Msolpyr$ and $v_{\rm
  cl} = 2000\,\kmps$, after which the first SN explosion (of $10^{51}
\,{\rm ergs}$ energy and $10\,\Msol$ ejecta) occurs.

The hydrodynamic grid covers a cubic region of $\pm 16\,$pc extent centered
on the GMC clump and contains $512^{3}$ cells. The cluster wind is
injected as purely thermal energy within a radius of 0.375\,pc (6
cells).  A uniform heating rate of $10^{-26} {\rm erg\,s^{-1}}$ is
used, together with a cooling curve designed to give 3 stable thermal
phases at $\sim$10, $100$, and $8000$\,K \citep[see][for further
details]{Pittard:2011}. The average particle mass in the simulations
is temperature dependent and determined by a look-up table
\citep[see][]{Sutherland:2010}.  
The densest regions cool
to 1\,K, which is the imposed temperature floor.  Our simulations do
not include gravity, thermal conduction or magnetic fields.

\begin{figure}[!t]
\begin{center}
\includegraphics[width=135mm]{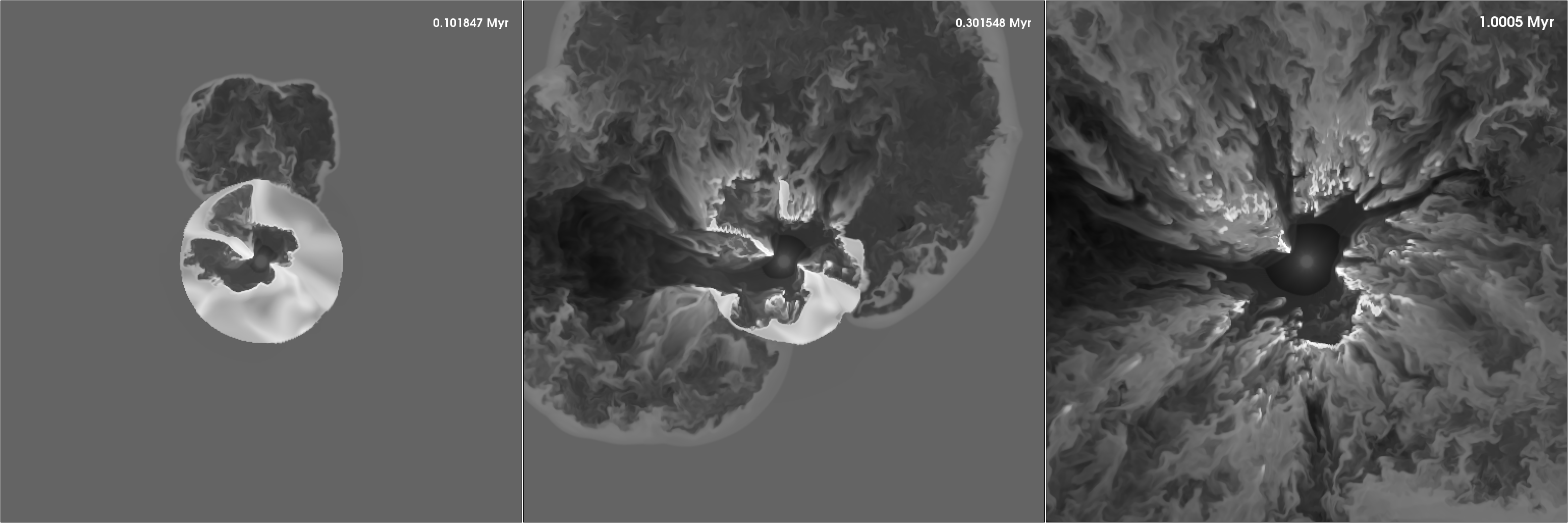}
\caption{Density slices through the 3D simulation at $t = 0.1$,
  0.3 and 1.0\,Myr. White is high density and black is lower density.} 
\label{fig:earlyevol}
\end{center}
\end{figure}

\section{Results}
The cluster wind creates a high-temperature bubble within the GMC clump
which expands most rapidly into regions of lower density and
pressure. Fig.~\ref{fig:earlyevol} shows the early evolution of this
process.  The cluster wind breaks out of the GMC clump wherever it can
find channels of low resistance. This break out ejects fragments of
the shell of GMC clump material swept up by the winds. Clump material
along the walls of these channels is ablated into the flow so that the
clump gradually loses mass. A wide variety of densities and
temperatues exists throughout the clump, with each covering in excess
of 7 orders of magnitude. A strong reverse shock heats the cluster
wind near its source, while many weaker shocks exist within the flow
as the surrounding dense regions collimate its passage. Slow shocks
are transmitted into dense clump material, which is compressed by the
high pressure wind. Dense parts of the clump can shield and protect
less dense material in their ``shadow'', though the ability of the
hot, high pressure, gas to flow around denser objects mitigates this
effect to some extent.
 
As lower density material is gradually cleared out of the clump, the
dense clouds slowly become isolated from each other, and find
themselves exposed more forcefully to streams of hot, fast-flowing,
gas. The material ablated from the clouds forms distinct tails.  The
lower dynamic pressure of the RSG-dominated cluster wind between 4.0
and 4.2\,Myrs slightly reduces the hydrodynamic ablation rate of the
clump. This reprieve is short-lived, however, as the cluster wind
becomes significantly more powerful when the stars enter their WR
stage. We find that the GMC clump loses mass at a roughly constant
rate from $t=0.0$ up to the end of the WR stage ($t = 4.5$\,Myr), with the mass of
molecular material declining from $5000\,\Msol$ to $1800\,\Msol$ over this
period.

The reverse shock in the cluster wind is far from spherical. Its shape
is strongly influenced by the presence of nearby dense clouds around
which the cluster wind forms bowshocks. At $t = 4.5$\,Myr the reverse
shock has a typical radius of 2.5\,pc, with the nearest dense clouds about
1.5\,pc from the centre of the cluster.

\begin{figure}[]
\begin{center}
\includegraphics[width=135mm]{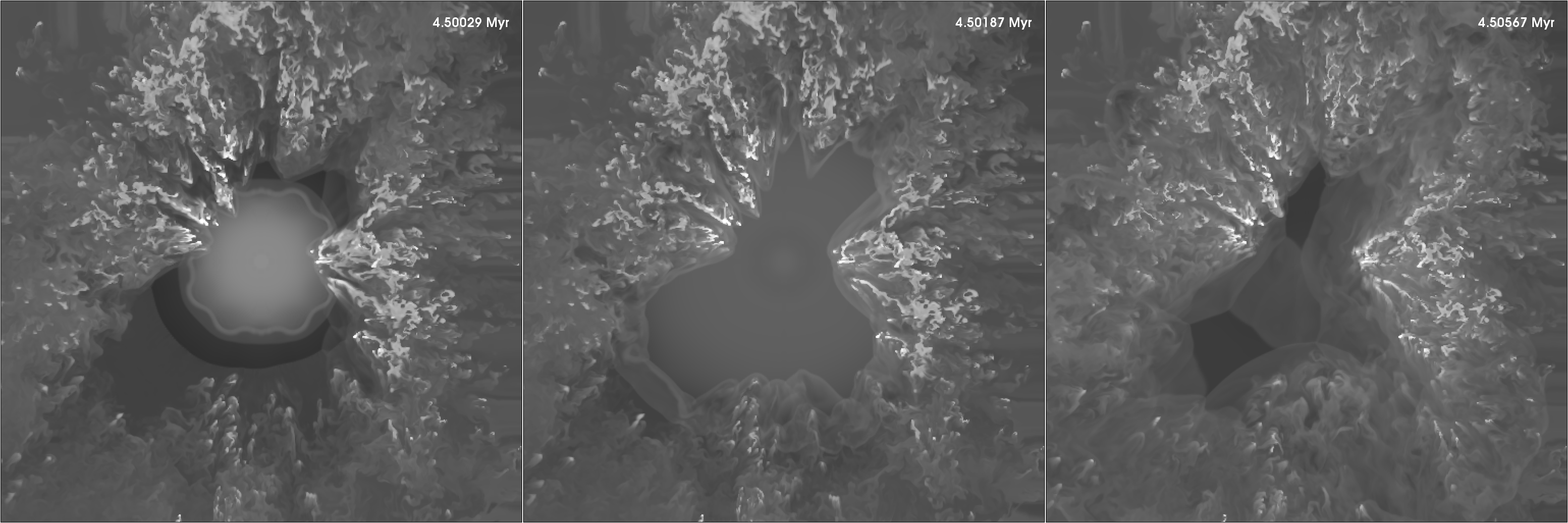}
\caption{Density slices through the 3D simulation at 960, 2540 and
  6340\,yrs after the first star explodes as a supernova. White is high density and black is lower density.} 
\label{fig:snexplosion}
\end{center}
\end{figure}

Fig.~\ref{fig:snexplosion} shows the time evolution of a density slice
through the cluster shortly after a SN explosion (occuring at
$t=4.49933$\,Myr) which deposits $10\,\Msol$ of ejecta and $10^{51}$\,erg of
energy into the surroundings. The SN ejecta sweeps up the
WR-dominated cluster wind into a thick, adiabatic shell. This shell
propagates at high speed through the lower density regions surrounding
the explosion, and refracts around the dense clouds which it
encounters. Transmitted shocks move into the dense clouds while
reflected shocks move backwards towards the explosion site and
converge in a highly asymmetic fashion due to the
azimuthally dependent distribution of the nearest dense clouds. The
early evolution of the supernova remnant (SNR) is hence very different
to the standard, spherically symmetric, picture. Our simulations
currently end at $t = 4.75$\,Myr.  The passage of the SNR appears to
have had very little immediate effect on the dense clouds.

\section{Conclusions and Future work}
Our preliminary analysis reveals that stellar wind feedback in massive
stellar clusters is a complex process which is strongly affected by
the inhomogeneity of the gas within the remains of the cluster-forming GMC
clump. The resulting structures and their evolution are far removed
from the results of simple spherically symmetric models. Hot, high
speed gas flows away from the cluster in low-density channels opened
up by the stellar winds. Mass is loaded into these flows from the
ablation of dense clouds embedded within them and from material
stripped from the dense gas which confines and directs the flows.  The
rate of ablation appears to be roughly constant during the first few
Myr of the cluster evolution. About half the initial molecular mass
remains by the time the first SN occurs (at 4.5\,Myr in our model). The
high porosity of the dense molecular material at this stage allows the
SN blast to rip through the cluster in a largely unimpeded fashion,
with the forward shock refracting around dense inhomogeneities. The
passage of the shock appears to have little immediate effect on the
remaining dense clouds.

A detailed analysis of our results is currently underway. However, it
would seem that stellar winds, like ionizing radiation fields
\citep{Dale:2011}, couple relatively weakly to the dense gas in such
environments. In a similar vein, the disruption of molecular clouds by
supernovae may also be weaker than previously anticipated.

Future work will examine massive star feedback on a variety of GMC
clump sizes, and will include the effects of magnetic fields, thermal
conduction and gravity.

\acknowledgements JMP would like to thank The Royal Society for 
funding a University Research Fellowship.


\begin{thebibliography}{}
\setlength\itemsep{0cm}

\bibitem[Bodenheimer, Tenorio-Tagle \& Yorke (1979)]{Bodenheimer:1979}
Bodenheimer, P., Tenorio-Tagle, G., Yorke, H. W.\ 1979, ApJ, 233, 85

\bibitem[Bruhweiler et al. (2010)]{Bruhweiler:2010}
Bruhweiler, F. C., Freire Ferrero, R., Bourdin, M. O., Gull, T. R.\ 2010, ApJ, 719, 1872

\bibitem[Castor, McCray \& Weaver (1975)]{Castor:1975}
Castor, J., McCray, R., Weaver, R.\ 1975, ApJ, 200, L107 

\bibitem[Chevalier \& Clegg (1985)]{Chevalier:1985}
Chevalier, R. A., Clegg, A. W.\ 1985, Nature, 317, 44

\bibitem[Dale \& Bonnell (2011)]{Dale:2011}
Dale, J. E., Bonnell, I.\ 2011, MNRAS, 414, 321

\bibitem[Domingo-Santamar\'{i}a \& Torres (2006)]{Domingo-Santamaria:2006}
Domingo-Santamar\'{i}a, E., Torres, D. F.\ 2006, A\&A, 448, 613

\bibitem[Draine \& Woods (1991)]{Draine:1991}
Draine, B. T., Woods, D. T.\ 1991, ApJ, 383, 621

\bibitem[Dyson \& de Vries(1972)]{Dyson:1972}
Dyson, J. E., de Vries, J.\ 1972, A\&A, 20, 223

\bibitem[Efstathiou (2000)]{Efstathiou:2000}
Efstathiou, G.\ 2000, MNRAS, 317, 697

\bibitem[Freyer, Hensler \& Yorke (2003)]{Freyer:2003}
Freyer, T., Hensler, G., Yorke H. W.\ 2003, ApJ, 594, 888

\bibitem[Harper-Clark \& Murray (2009)]{Harper-Clark:2009}
Harper-Clark, E., Murray, N.\ 2009, ApJ, 693, 1696

\bibitem[Krumholz et al. (2010)]{Krumholz:2010}
Krumholz, M. R., Cunningham, A. J., Klein, R. I., McKee, C. F.\ 2010, ApJ, 713, 1120

\bibitem[Lopez et al. (2011)]{Lopez:2011}
Lopez, L. A., Krumholz, M. R., Bolatto, A. D., Prochaska, J. X., Ramirez-Ruiz, E.\ 2011, ApJ, 731, 91

\bibitem[Mac Low et al. (2007)]{MacLow:2007}
Mac Low, M.-M., Toraskar, J., Oishi, J. S., Abel, T.\ 2007, ApJ, 668, 980

\bibitem[Matzner (2002)]{Matzner:2002}
Matzner, C. D.\ 2002, ApJ, 566, 302

\bibitem[Mazurek (1980)]{Mazurek:1980}
Mazurek, T. J.\ 1980, A\&A, 90, 65

\bibitem[McKee \& Ostriker(1977)]{McKee:1977}
McKee, C. F., Ostriker, J. P.\ 1977, ApJ, 218, 148

\bibitem[Mellema et al. (2006)]{Mellema:2006}
Mellema, G., Arthur, S. J., Henney, W. J., Iliev, I. T., Shapiro, P. R.\ 2006, ApJ, 647, 397

\bibitem[Peters et al. (2010)]{Peters:2010}
Peters, T., Klessen R. S., Mac Low, M.-M., Banerjee R.\ 2010, ApJ, 725, 134

\bibitem[Pittard (2011)]{Pittard:2011}
Pittard, J. M.\ 2011, MNRAS, 411, L41

\bibitem[Reyes-Iturbide et al. (2009)]{Reyes-Iturbide:2009}
Reyes-Iturbide, J., Vel\,{a}zquez, P. F., Rosado, M., Rodr\,{i}guez-Gonz\,{a}lez,
A., Gonz\,{a}lez, R. F., Esquivel, A.\ 2009, MNRAS, 394, 1009

\bibitem[Rodr\'{i}guez-Gonz\'{a}lez et al. (2008)]{Rodriguez-Gonzalez:2008}
Rodr\'{i}guez-Gonz\'{a}lez, A., Esquivel, A., Raga, A. C., Cant\'{o},
J.\ 2008, ApJ, 684, 1384

\bibitem[Sutherland (2010)]{Sutherland:2010}
Sutherland, R. S.\ 2010, Ap\&SS, 327, 173

\bibitem[Tenorio-Tagle (1979)]{Tenorio-Tagle:1979}
Tenorio-Tagle, G.\ 1979, A\&A, 71, 59

\bibitem[Tenorio-Tagle, Bodenheimer \& Yorke (1985)]{Tenorio-Tagle:1985}
Tenorio-Tagle, G., Bodenheimer, P., Yorke, H. W.\ 1985, A\&A, 145, 70

\bibitem[Tsivilev et al. (2002)]{Tsivilev:2002}
Tsivilev, A. P., Poppi, S., Cortiglioni, S., Palumbo, G. G. C.,
Orsini, M., Maccaferri, G.\ 2002, New Astr., 7, 449

\bibitem[Toal\'{a} \& Arthur (2011)]{Toala:2011}
Toal\'{a}, J. A., Arthur, S. J.\ 2011, ApJ, 737, 100

\bibitem[Townsley et al. (2003)]{Townsley:2003}
Townsley, L. K., Feigelson, E. D., Montmerle, T., Broos, P. S., Chu, Y.-H., Garmire, G. P.\ 2003, ApJ, 593, 874

\bibitem[Townsley et al. (2006)]{Townsley:2006}
Townsley, L. K., Broos, P. S., Feigelson, E. D., Garmire, G. P., Getman, K. V.\ 2006, AJ, 131, 2140

\bibitem[V\'{a}zquez-Semadeni et al. (2008)]{Vazquez-Semadeni:2008}
V\'{a}zquez-Semadeni, E., Gonz\'{a}lez, R. F., Ballesteros-Paredes, J., Gazol, A., Kim, J.\ 2008, MNRAS 390, 769

\bibitem[Wang et al. (2010)]{Wang:2010}
Wang, P., Li, Z.-Y., Abel, T., Nakamura, F.\ 2010, ApJ, 709, 27

\bibitem[Weaver et al. (1977)]{Weaver:1977}
Weaver, R., McCray, R., Castor, J., Shapiro, P., Moore, R.\ 1977, ApJ, 218, 377

\bibitem[White \& Rees(1978)]{White:1978}
White, S. D. M., Rees, M. J.\ 1978, MNRAS, 183, 341

\bibitem[Whitworth (1979)]{Whitworth:1979}
Whitworth, A.\ 1979, MNRAS, 186, 59

\bibitem[Yorke et al. (1989)]{Yorke:1989}
Yorke, H. W., Tenorio-Tagle, G., Bodenheimer, P., Rozyczka, M.\ 1989, A\&A, 216, 207

\end{thebibliography}
\end{document}